# HYDRODYNAMIC OF SURGE WAVE ENERGY CONVERTERS

ABBAS YEGANEH-BAKHTIARY (1), MILAD REZAEE (2)
*(1): School of Civil Engineering, IUST, Narmak, Tehran,Iran*
*(2): School of Civil Engineering, IUST, Narmak, Tehran,Iran*

Ocean wave energy is a new renewable energy resource which is going to become one of the reliable and alternative resources for fossil fuels during recent decades. The majority of studies have focused on extract wave energy at an effective rate; whilst, there are a few studies to explore the hydrodynamic of wave and surge wave energy converters. In this study a 2D numerical model based on RANS equations is closured with SST turbulence model employed to simulate the hydrodynamic of the flap type wave energy devices. The results indicate that a partially standing wave induced by the interaction of incident and reflected waves occurs in front of device due to its rotating movements.

**Keywords: Wave energy, Wave energy converter, Steady streaming, Scour, FLUENT**

## INTRODUCTION

Renewable energy of wind generated waves is a new energy resources which has the potential to become a reliable alternative resource for fossil fuels. The majority of studies in this regard have focused on extract wave energy at an effective rate whilst there are very limited studies to determine the environmental impacts of wave energy devices. In a study, the effect of turbulent and vortex shedding induced by a wave energy device on other devices and on the device's environment has been evaluated Amoudry *et al*. [1]. Williams and Esteves emphasized the sensitivity of the shoreline to the characteristics of the incoming waves. Millar *et al*. assessed the impact of the changes in a wave farm due to Wave Hub off the north coast of Cornwall [2].

Wind waves propagate in water body and if there would be no limitation in fetch and duration, resultant waves will carry enough energy to move flap type wave energy converters as one of the most effective devices which place in intermediate depths e.g. 10 to 20 meter depths. Folley *et al*. [3] showed that small seabed-mounted wave energy devices receive 5 to 10 percent less wave power in compare to offshore devices; however extracted energy in such devices are lower in cost due to installation and maintenance costs. In another study Folley *et al*.[4]indicated that in order to maximize the power captured, the flap must be tuned to the incoming wave frequency. Also the motion angel should be less than 30 degree.

Numerous studies have been carried out on the efficiency of wave energy converters or modeling the motion of the device, however there is no study to analyze or model the effect of moving flap on incident wave while like as other marine structures such as vertical walls, it is crucial to understand the impact of the structure on the incident wave, the velocity field in front of the flap and the interaction of reflected and incoming waves.

Standing wave occurs as the result of the superposition of incident waves and reflected waves, when a series of waves confront an obstruct such as a vertical wall .To the best knowledge of the authors little information is available for the hydrodynamic and velocity field in front of the surge wave energy devices. However the concept of standing and partially waves in front of vertical and inclined walls is not a new issue and there is a numerous studies in this field. The effect of standing waves was observed by Carter *et al.* [5]. They indicated that the resultant standing waves generate a velocity field which named steady streaming that it consists two essential up and bottom recirculation cells. Hajivalie *et al.* [6] employed a numerical model to simulate standing waves in front of a vertical wall. Yeganeh-Bakhtiary *et al.* [7] developed a numerical model based upon RANS equations model to simulate partially standing waves and the resultant hydrodynamics in front of a vertical wall.

Our main object in this study is investigation of the influence of a moving flap in a surge wave energy converter on wave field and the resultant standing waves in front of that. A numerical model which solves two-dimensional RANS equations has been used. VOF technique was applied in order to calculate the free surface and SST $k-\omega$ turbulence model was used to estimate the turbulence stress.

## GOVERNING EQUATIONS

To simulate the flow in front of surge WEC, unsteady two-dimensional NS equations are applied. Since the flow is turbulent an appropriate turbulence model which is SST turbulence model is closured with the flow model. Governing equations which embracing momentum, continuity and SST equations are introduced in two-dimensional coordinates as follow:

$$\frac{\partial U}{\partial x} + \frac{\partial W}{\partial z} = 0 \tag{1}$$

$$\frac{\partial U}{\partial t} + U\frac{\partial U}{\partial x} + W\frac{\partial U}{\partial z} = -\frac{1}{\rho}\frac{\partial P}{\partial x} + \frac{\partial}{\partial x}\left(2\Gamma\frac{\partial U}{\partial x}\right) + \frac{\partial}{\partial z}\left(\Gamma\left(\frac{\partial U}{\partial z} + \frac{\partial W}{\partial x}\right)\right) \tag{2}$$

$$\frac{\partial W}{\partial t} + U\frac{\partial W}{\partial x} + W\frac{\partial W}{\partial z} = -\frac{1}{\rho}\frac{\partial P}{\partial z} - g + \frac{\partial}{\partial x}\left\{\Gamma\left(\frac{\partial U}{\partial z} + \frac{\partial W}{\partial x}\right)\right\} + \frac{\partial}{\partial z}\left(2\Gamma\frac{\partial W}{\partial z}\right) \tag{3}$$

in which $x$ and $z$ are horizontal and vertical coordinates respectively, $U$ and $W$ are mean velocity components of flow in $x$ and $z$ directions, $P$ is the mean pressure, $g$ is the gravity acceleration, $\rho$ is the fluid density. Based on [8], [9] two-dimensional SST equations are:

$$v_t = \frac{\alpha_1 k}{\max(\alpha_1 \omega, SF_2)} \tag{4}$$

$$\frac{\partial k}{\partial t} + U_i \frac{\partial k}{\partial x} = P_k - \beta^* k\omega + \frac{\partial}{\partial x}\left[\left(v + \sigma_k v_t\right)\frac{\partial k}{\partial x_i}\right] \tag{5}$$

$$\frac{\partial \omega}{\partial t} + U_j \frac{\partial \omega}{\partial x_j} = \alpha S^2 - \beta \omega^2 + \frac{\partial}{\partial x_j}\left[(\nu + \sigma_\omega \nu_t)\frac{\partial \omega}{\partial x_j}\right]$$
$$+ 2(1-F_1)\sigma_{\omega 2}\frac{1}{\omega}\frac{\partial k}{\partial x_i}\frac{\partial \omega}{\partial x_i} \tag{6}$$

$$F_2 = \tanh\left[\left[\max\left(\frac{2\sqrt{k}}{\beta^* \omega z}, \frac{500\nu}{z^2 \omega}\right)\right]^2\right] \tag{7}$$

$$P_k = \min\left(\tau_{ij}\frac{\partial U_i}{\partial x_j}, 10\beta^* k \omega\right) \tag{8}$$

$$F_1 = \tanh\left\{\left\{\min\left[\max\left(\frac{\sqrt{k}}{\beta^* \omega z}, \frac{500\nu}{z^2 \omega}\right), \frac{4\sigma_{\omega 2} k}{CD_{k\omega} z^2}\right]\right\}^4\right\} \tag{9}$$

$$CD_{k\omega} = \max\left(2\rho\sigma_{\omega 2}\frac{1}{\omega}\frac{\partial k}{\partial x_i}\frac{\partial \omega}{\partial x_i}, 10^{-10}\right) \tag{10}$$

Folley *et al.* [4] suggested to model the dynamics of seabed-mounted bottom hinged wave energy converters by a single degree of freedom model which would be [10]:

$$T_w(t) = (I + I_\infty)\ddot{\theta} + \int_0^t \dot{\theta}(\tau)K(t-\tau)d\tau + k_p\theta + \Lambda\dot{\theta} + B_\nu|\dot{\theta}|\dot{\theta} \tag{12}$$

In which $T_w(t)$ is the wave torque in time *t*, *I* is the body moment of inertia, $I_\infty$ is added moment of inertia in unlimited frequencies, $\theta$ is the rotation angle, *K(t)* is the response function, $k_p$ is rotational stiffness, $\Lambda$ is damping coefficient and $B_\nu$ is viscose damping coefficient. Folley *et al.* [4] suggested that the rotation of flap constrained to a maximum angle of $\theta_{max} \leq 30°$.

**NUMERICAL DOMAIN**

A numerical model schema was carried out to simulate the hydrodynamic in front of a surge wave energy converter. The computational domain was meshed by a composite of uniform structure and unstructured mesh. The size of grids in air phase was greater than the water phase and also a finer grid was used near boundaries. Since the wave energy device is a moving wall, which rotates around its hinged point in seabed, dynamic mesh system has been used. Figure 1 shows the mesh structure and the computational domain of the simulation. It is shown that mesh size of air phase is coarser than water phase and it is finer in boundary layers.

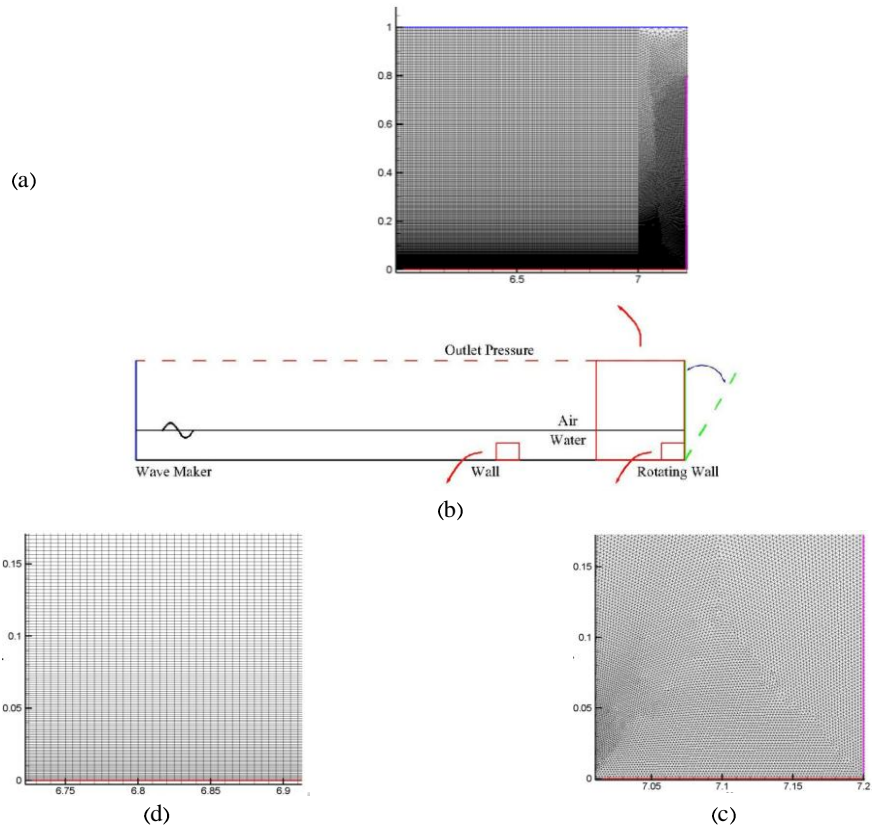

Figure 1. Mesh structure in current study

In order to discrete the governing equations, a second order upwind schema was applied. The simulation was carried out by Fluent software. Stability of numerical schema was controlled mainly by time step or local Courant number.

Table 1. The incident wave characteristics based on Xie [11]

| Test No. | H (m) | T (sec) | h (m) | L (m) | $H/gT^2$ | $d/gT^2$ |
|---|---|---|---|---|---|---|
| 1 | 0.050 | 2.41 | 0.30 – 0.45 | 4.0 | 0.0008 | 0.007 |
| 2 | 0.065 | 1.53 | 0.30 – 0.45 | 2.4 | 0.0028 | 0.019 |
| 3 | 0.060 | 1.86 | 0.30 – 0.45 | 3.0 | 0.0017 | 0.013 |

Figure 2 presents a cooperation between numerical and experimental maximum horizontal velocity namely near the first node of standing waves at the halfway of nodes and anti-nodes. As it is shown, current numerical model's results graph the agreement is quite well.
In current study, in order to validate the fluid model, we used experimental data of Xie [11] model to compare orbital velocities in standing wave condition for vertical wall. The wave flume that the experiment was conducted in, was a 38.0 m long, 0.8 m wide and 0.6 m deep. At the beginning, the water depth was 45.0 cm and it transformed to 30.0 cm with a 1:30 slope in front of the breakwater. The incident waves varied from 5.0 to 9.0 cm and the wave period varied from 1.17 to 3.56 sec. Table 1 gives the information of three different experiments of this study which are discussed here

**RESULTS AND DISCUSSION**

Computational domain, mesh structure and boundary and initial conditions have been discussed in previous sections. Figure 3 shows the partially standing wave field during a wave period in front of surge-WEC. It is shown that the greater angle of the wall results the increase in wave steepness and consequently the increment in wave non linearity.

The stream lines in a wave period are shown in Fig. 4. In time t=0.0, the formation of streamlines are like as theoretical formation in standing wave condition. Over the time, with inclination of the wall in front of incoming waves, streamlines get longer which shows the moving of the water particles toward the device or in other words the flow in front of the device. Thus during a wave period there would be a flow in front of surge wave energy device and hence it does not form a completely standing waves due to the moving of the flap. The flow cause non linearity terms in wave field and formation of partially standing waves. In addition the orbital motion of water particles and the created flow leads to a displacement of nodes and antinodes' location.

In Fig. 5 the vertical and horizontal velocities in place of nodes and antinodes in front of surge-WEC is shown. The figure shows that the velocity profiles in these points are inversed. Approximately in 0.01 meter of bed, the horizontal velocity profile experienced the change in the sign for both node and antinodes, which means if the velocity was positive it changed to negative during the 0.01 meter form seabed. In 0.12 meter form the bed, horizontal velocity profile crosses the vertical velocity profile and in 0.25 meter the direction of horizontal velocities changed again.

**CONCLUSION**

Investigation of the impact of surge wave energy converters on wave field is one the issues that despite its importance in the design of these structures, has received less consideration in compare to other issue in related to ocean wave energy. In order to study the hydrodynamics of wave-surge wave energy converters including the velocity field and recirculation flows, a two-phase flow framework is used to study the impact of moving flap of surge-WEC on wave field. The numerical model is based on two-dimensional RANS equations in closure with SST turbulence model and the two-phase VOF technique. The method was verified against the experimental data of a vertical wall of Xie (1981). The comparison between model and experimental data of vertical wall showed that the model is suitable to simulate the impact of an obstacle on a wave field thus because of the shortcoming of data for wave-surge WECs interaction, we concluded that the model can simulate the currents in front of these wave energy devices that work like as a moving wall in front of incoming waves.

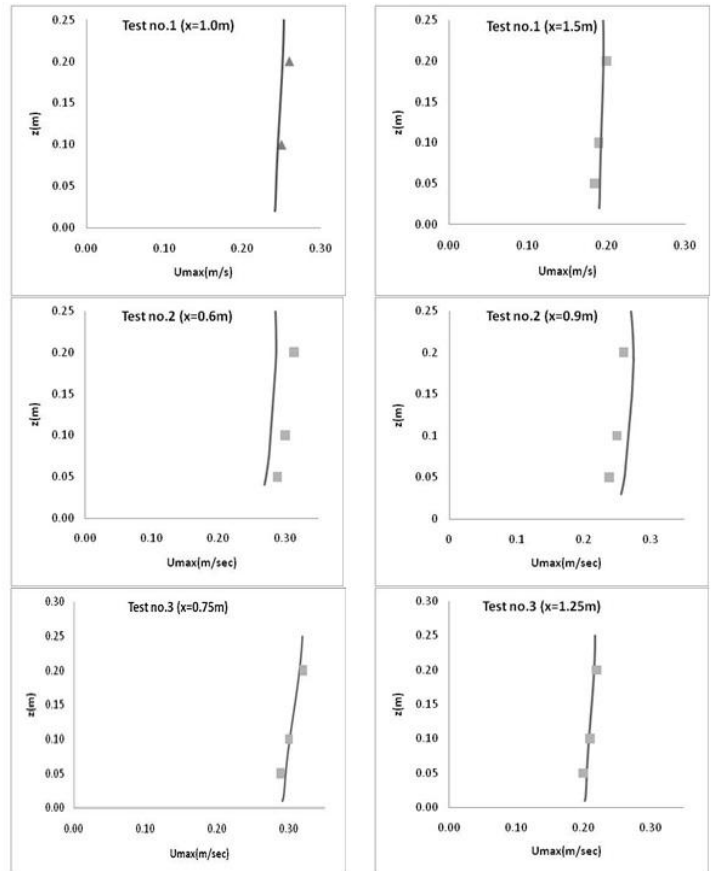

Figure 2. Comparison between numerical and experimental maximum horizontal velocity

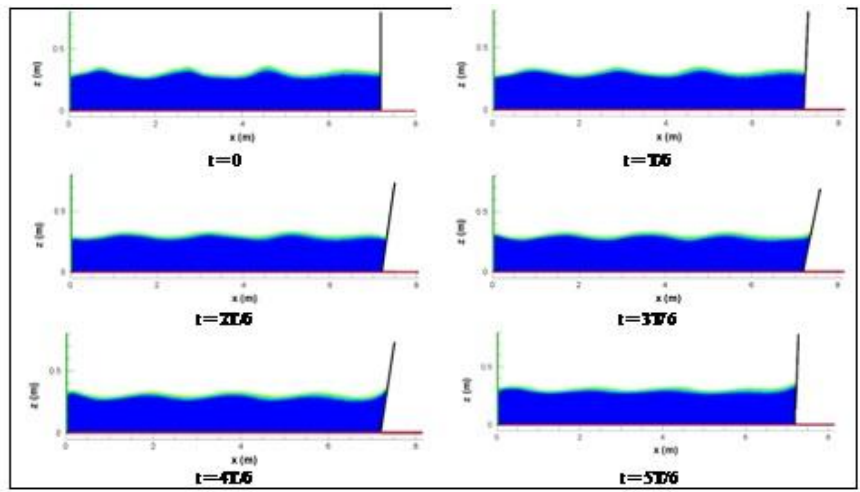

Figure 3. Partially standing wave field during a wave period in front of surge-WEC

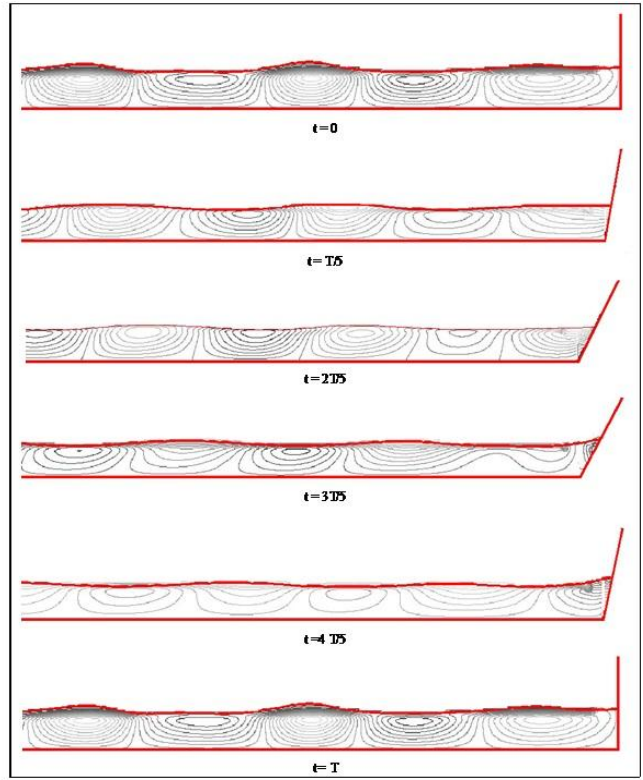

Figure 4. Streamlines during a wave period in front of surge-WEC

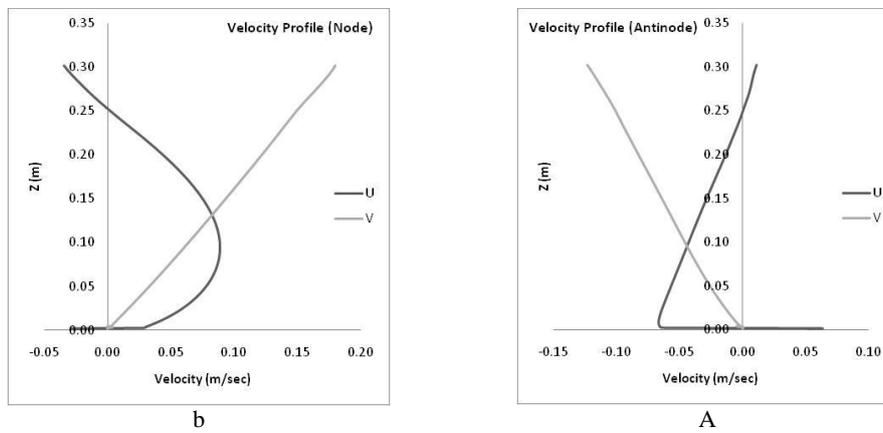

Figure 5. Vertical and horizontal velocities in place of nodes and antinodes in front of surge-WEC

The results showed that the rotation of the wall which its result is absorption a part of incoming energy and reduction in reflected wave's energy and velocity, the reflected waves have less height and different frequency in compare to incoming waves. Thus we have partially standing waves in front of these devices.